\documentclass{article}
\usepackage{graphicx} 
\usepackage{algorithm}
\usepackage{algpseudocode}
\usepackage{algorithmicx}
\usepackage{listings}
\usepackage{color}
\usepackage{amsmath}  
\usepackage{graphicx} 
\usepackage{amssymb} 
\usepackage{algpseudocode}
\usepackage{algorithm}
\usepackage{comment}
\usepackage{hyperref}
\usepackage{orcidlink}
\usepackage{pgfplots}
\usepackage[english]{babel}
\usepackage{float}
\floatstyle{plaintop}
\restylefloat{table}
\usepackage{pythonhighlight}
\usepackage{lscape}

\usepackage{graphicx}
\usepackage{epstopdf}
\usepackage{algorithm}
\usepackage{algpseudocode}
\AppendGraphicsExtensions{.gif}

\usepackage{booktabs}
\usepackage[utf8]{inputenc}
\usepackage[T1]{fontenc}
\usepackage{tabularx}
\usepackage{datetime}

\usepackage{placeins}
\usepackage{comment}
\usepackage{breakcites}
\usepackage{tabto}
\usepackage{listings}
\usepackage{titlepic}
\usepackage{graphicx}
\graphicspath{ {./figures/} }
\usepackage{multirow}

 \date{}

\makeatletter
\providecommand{\institute}[1]{
  \apptocmd{\@author}{\end{tabular}
    \par
    \begin{tabular}[t]{c}
    #1}{}{}
}
\makeatother

\title{Energy Consumption Framework and Analysis of Post-Quantum Key-Generation on Embedded Devices}

\author{J Cameron Patterson, William J Buchanan, Callum Turino}
\institute{Blockpass ID Lab, Edinburgh Napier University}

\pgfplotsset{compat=1.18}  

\begin{document}

\maketitle

\begin{abstract}
The emergence of quantum computing and Shor's algorithm necessitates an imminent shift from current public key cryptography techniques to post-quantum robust techniques.  NIST has responded by standardising Post-Quantum Cryptography (PQC) algorithms, with ML-KEM (FIPS-203) slated to replace ECDH (Elliptic Curve Diffie-Hellman) for key exchange.  A key practical concern for PQC adoption is energy consumption.  This paper introduces a new framework for measuring the PQC energy consumption on a Raspberry Pi when performing key generation. The framework uses both available traditional methods and the newly standardised ML-KEM algorithm via the commonly utilised OpenSSL library.

\end{abstract}

\section{Introduction} \label{Introduction}

Ensuring data confidentiality relies heavily on cryptographic techniques. Whilst single-key symmetric encryption such as AES-256 \cite{fips197} offers efficiency for bulk data encryption, the secure establishment and rotation of the shared keys used is critical to the success of maintaining confidentiality. A primary challenge of encryption is that keys require sharing between parties, but that the channel between them may be subject to interception and eavesdropping. 
Many of the activities that are carried out online: from Internet banking, to commerce sites or reading the news, all now typically use secure HTTPS and encryption. 
If a key used for encryption were to become known, then in some applications there may be severe consequences as it could potentially be used to impersonate either party in the transaction or to decrypt that data.
Given the vast scale and dynamic combination of users and sites, it is not practical to pre-generate and distribute all cryptographic key combinations in advance, 
so the keys must be negotiated ad hoc: as they become required online.
It is therefore critically important to ensure that encryption keys are negotiated securely and kept protected from potential adversaries.

Fortunately, cryptographic keys are typically generated and negotiated using asymmetric cryptography techniques first disclosed to the academic community by Diffie and Hellman \cite{diffie1976new}.
Here, each party has its own key-pair comprising their own private and public keys, and a mathematical technique is used to combine the other party's public key with a local secret private key to negotiate a secure channel in which only those two parties can participate. 
Assuming a strong method, this addresses the issues of eavesdropping and confidentiality, and also addresses a malicious actor attempting to interpose themselves into communications. 

\subsection{Key Generation with Hybrid Encryption}
Whilst this asymmetric method can be used for data encryption, it is typically far less efficient (in time and energy resources) than for communications protected with symmetric encryption \cite{boneh2003comparative}.
The asymmetric channel's typical main purpose therefore becomes the secure sharing of the symmetric algorithm's key material, with the bulk data then conveyed symmetrically encrypted with this key. 
This standard approach to modern cryptography is known as `hybrid encryption' \cite{turner2001hybrid}, trading-off the characteristics of each technique for reasons of convenience, efficiency and confidentiality.

Asymmetric key generation techniques are a critical part of current hybrid cryptography, with popular methods including RSA \cite{RSA1978method} and Elliptic Curve Cryptography (ECC) \cite{koblitz1987elliptic,miller1986use}. 
If these methods were to be fundamentally compromised, then access to the negotiated secret symmetric keys could be revealed, with the consequence of data decryption becoming trivial. 

\subsection{Quantum Computing Exploits}
For some time it has been known that the mathematical principles underpinning RSA and ECC's security, whilst computationally intensive and difficult for a classical computer-based attack, are potentially readily solvable with quantum computing hardware.
Shor's algorithm \cite{shor1994} is notable in this respect, as it is able to factorise large numbers extremely efficiently. 
This development necessitates the creation of alternative techniques for secure key exchange in a `post-quantum' world, particularly as encrypted data could be stored and recovered later using these methods.
In response to this threat, the National Institute of Standards and Technology (NIST) has been a leader, initiating a competition in 2017 to identify promising post-quantum cryptography (PQC) methods \cite{nist_pqc_csrc}. These techniques rely on differing mathematical problems, believed to be resistant to both known classical and quantum computing algorithms.

\subsection{Aim of the Paper}
The key aim of this paper is to analyse and compare the energy efficiency of standard pre- and post-quantum NIST cryptographic key generation algorithms, as implemented in the OpenSSL 3.5 library released in April 2025 \cite{openssl35}.
It differs from other evaluations, which typically concentrate on time performance rather than energy performance e.g. `openssl speed' to give tables such as in \cite{asecuritysite_13721} and from other work in the energy recording area such as Tasopoulos et al. \cite{TLS13ResourceConstrained}.
It creates a framework for the readily-available Raspberry Pi and makes use of accurate,  commodity USB energy meters.

The Raspberry Pi is a popular single-board computer platform increasingly utilised in industrial process applications \cite{bahrak2020raspberry} where controllers are typically embedded with long operational lifetimes and where energy constraints can be a defining factor in design and implementation.

\section{Related Literature} \label{LiteratureReview}

\subsection{Post-Quantum Cryptography (PQC)}

The imminent emergence of quantum computing is fast becoming a problem in reality not only for cryptographers but also for users of cryptography.
What was once purely a theoretical problem relegated to the `long grass' of the future has emerged as a pressing concern within the 2020s, as Aydeger et al. outline \cite{PQCTransition}. 
The potential impact of quantum computing on cryptography poses a significant threat to the existing hegemony of cryptographic standards. 
These techniques, including RSA \cite{RSA1978method} (from 1978), DES/3DES \cite{DES,3DES} (1977/1995), and others more recently including AES \cite{fips197} (2001) and Elliptic Curve Cryptography (ECC) whose applicability is outlined in papers such as Koblitz \cite{koblitz1987elliptic} and Miller \cite{miller1986use} from the late 1980s, have proven to be the bedrock of the practical protection of communications over the last decades \cite{PQCTransition}.

\subsubsection{In the Classical Space}
The main algorithms in contemporary use are not currently known to have fundamental flaws that can be exploited by attacks based on classical computing. 
Whilst classical computing capabilities have risen exponentially, this increase has been largely driven by Moore's Law \cite{MooreSchaller} and the self-governed targets of the semiconductor industry, as outlined in reports such as the IEEE International Roadmap for Devices and Systems \cite{irds_metrology_2023}. 
Specialised computing branches, including massively parallel GPUs, ASICs, and FPGAs, have achieved performance gains that outpaced Moore's Law for general-purpose computing hardware, further contributing to the potential for stronger cryptographic algorithm attacks. 
Despite these improvements, the underlying mathematical problems on which cryptographic techniques rely have remained largely intact, with key sizes increasing to meet user expectations for security, and the required window of protection. 
One such mathematical characteristic is the computational hardness of large-number factorisation, a principle that protects both key exchange and the integrity of digital signatures. 
By using appropriately chosen cryptographic parameters, data protected by these traditional methods is expected to remain secure for decades, centuries, or even millennia, due to the time and energy required to mount a successful attack using classical computing methodologies.

\subsubsection{The Rise of the `Quantum' Machines}
The general availability of powerful quantum computers would fundamentally alter the capability to attack existing cryptography standards, and significant progress has been made in theoretical attack vectors and the production of small-scale quantum computers. 
Although progress achieved may currently seem limited, technological improvements often occur on a logarithmic scale (c.f., Moore's Law \cite{MooreSchaller}), with seemingly slow initial progress rapidly transforming into significant advancement. 
Algorithms such as Shor \cite{shor1994} exemplify this threat, reducing what was once a logarithmically difficult factorisation problem for classical devices into one that can be solved relatively easily using sufficiently large quantum computers. Quantum computing has the capacity to perform calculations across many states simultaneously, breaking the logarithmic protection of the techniques commonly deployed including ECC and RSA. 
Widespread quantum computing availability in the near future poses a particular concern, as adversaries may already be engaging in `Store Now, Decrypt Later' (SNDL)  activities in anticipation.
This is clearly a concern for sectors where long-term data confidentiality is crucial, such as the protection of proprietary information in industry and `top-secret' government activities.

\subsubsection{Standardisation and Governmental Response}
The risks highlighted by quantum computing have focused the attention of governments, industry, and the research community, prompting a wide range of response activities. 
The research presented in this paper was selected to contribute to the development and validation of new post-quantum cryptography (PQC) techniques and standards, specifically examining the practical reality of energy use.
PQC methods currently appear to be more resilient to both classical and potential quantum computing attacks than existing key-generation algorithms including RSA and ECC \cite{nist_pqc_csrc}. 
Governments worldwide are establishing timelines for the transition away from quantum-vulnerable algorithms to quantum-resistant techniques. 
Australia's approach is among the most ambitious, with a target to complete the transition by 2030 \cite{cyber_gov_au_cryptography}.
Many other countries and administrative domains are also setting ambitious deadlines in the 2030s, including the UK \cite{ncsc_pq_timelines}, the US \cite{nistir8547}, and the EU (through individual agencies in member states) \cite{eu_pqc_joint_statement_2024,eu_pqc_recommendation_2024}.

\subsection{Vulnerabilities Exposed By Quantum Computing}


Shor's algorithm, presented in 1994 \cite{shor1994}, poses a significant theoretical threat to traditional public-key cryptography, and is well-known as a critical issue which must be addressed. 
Shor offers a `polynomial-time quantum algorithm' for factorising large integers, a computationally hard problem for classical computers, and is the basis of many classical key exchange mechanisms. 
Although highly speculative, some industry estimates such as those from MITRE \cite{Mitre2025quantumEst}, suggest that attacks on RSA-2048 could become feasible by 2035.
It is probable that `Store Now, Decrypt Later' activities are already underway awaiting quantum computing's potential.
This highlights why the subject is receiving urgent attention in critical government, military, and commercial applications where confidentiality time-frames must be maintained beyond the prospective availability of quantum computing techniques.

Another significant quantum algorithm that threatens modern cryptography is Grover's algorithm: `A fast quantum mechanical algorithm for database search' \cite{grover1996} (1996). 
Grover's algorithm provides a quadratic speedup for searching unsorted databases (such as the AES keyspace), effectively halving the key length of symmetric algorithms like AES \cite{grover1996}, thus AES-256's effective security reduces to AES-128. 
Whilst this still provides a sufficiently-sized window for the foreseeable future, halving the effective number of bits weakens the encryption and leaves it more vulnerable should further advances be made in this area.

Despite their theoretical nature, the increasing feasibility of quantum computing has driven significant attention to Post-Quantum Cryptography (PQC) at regulatory and governmental levels. 
While the extension of existing algorithms has been explored with larger keys for example, the consensus is that practical RSA is fundamentally vulnerable to Shor's algorithm on a sufficiently powerful quantum computer. 
Simply increasing the key size is not a viable long-term solution with Bernstein et al. \cite{1TBRSA} suggesting impractical key sizes (e.g., 1TB) would be needed for post-quantum safety. 
Figure \ref{EnergyComparePlot} in Section \ref{Implementation} of this paper demonstrates the exponentially increasing resource demands (time and energy) for generating larger RSA key pairs (e.g., 4096-bit). 
Whilst the Grover algorithm's impact is considered less catastrophic compared with the potential of Shor's, the cryptographic community is also actively exploring ways to reinforce symmetric encryption security, and address other limitations of algorithms like AES \cite{newAES} after its 20+ years of practical use.

\subsection{NIST Standarization}

In 2022, the National Institute of Standards and Technology (NIST) finalised its first set of quantum-resistant cryptographic algorithms \cite{nist_pqc_csrc} following a standardisation process initiated in 2017. 
This competition aimed to identify cryptographic methods resilient to both classical and quantum computer attacks. 
The selected algorithms are published as Federal Information Processing Standards (FIPS), establishing them as de-facto standards suitable for production use \cite{nist_pq_standards_FIPS203-205}. 
NIST's FIPS standards are widely recognised benchmarks for cryptography, making NIST's PQC selections likely to be adopted by industry and governments internationally.

This paper focuses primarily on FIPS 203 \cite{fips203}, which standardises ML-KEM (formerly Kyber). 
ML-KEM is a Key Exchange Mechanism (KEM) that securely establishes a session between two parties using asymmetric encryption, subsequently enabling a shared secret symmetric key to be agreed and used for the secure encryption of data transfers between the parties. 
It employs lattice-based cryptography, relying on the mathematical difficulty of lattice problems like Shortest Vector Problem (SVP) and Learning With Errors (LWE), believed to be hard to attack by both classical and quantum computers \cite{nist_pq_standards_FIPS203-205}. 
ML-KEM is designed to replace existing key-exchange protocols in applications e.g., for websites and setting up other secure channels. 


Whilst ML-KEM is currently the only finalised FIPS standard for post-quantum key exchange, in March 2025 NIST furthermore selected Hamming Quasi-Cyclic (HQC) \cite{nist_pqc_round4_status} as an alternative method to head towards standardisation. 
For this paper's experimentation, ML-KEM is used owing to its integration into the OpenSSL 3.5 production library \cite{openssl35}. The first batch of NIST PQC standards \cite{nist_pq_standards_FIPS203-205} also included two digital signature algorithms: FIPS 204 \cite{fips204}, standardising ML-DSA (formerly Dilithium); and FIPS 205 \cite{fips205}, detailing SLH-DSA (formerly SPHINCS+).



\subsubsection{Beyond the First Batch PQC standards - Round 4}

Many of the cryptographic techniques in the PQC process are based on lattice-based category methods \cite{nist_pqc_round4_status}, which unlike other techniques do not have a long history of use and validation, nor have formal proofs attached to them.
NIST therefore sought diverse algorithm candidates to mitigate risks associated with potential vulnerabilities in a specific category (specifically lattices), as noted on page 4 of the Round-4 NIST PQC status paper \cite{nist_pqc_round4_status}.

As part of the evaluation of these candidates, the Isogeny-based key exchange candidate SIKE was withdrawn after a flaw was discovered - see pp. 1, 8 and 16 of \cite{nist_pqc_round4_status}. 
Subsequently, NIST selected Hamming Quasi-Cyclic (HQC) per page 18 in the same document \cite{nist_pqc_round4_status}, a non-lattice-based key exchange protocol utilising `quasi-cyclic codes', for standardisation. 
This alternative will proceed towards formalisation with a designated name and FIPS standard. 
Having multiple standardised options also allows for easier adoption of alternatives based on the environment requirements of the deployment, such as bounds of working memory or key length. 
This desire for diverse characteristics across all areas is highlighted throughout the Round 4 status paper \cite{nist_pqc_round4_status}.

\subsection{ Energy Efficiency in PQC}
This section reviews studies on the energy efficiency of post-quantum cryptography (PQC) algorithms, especially within Transport Layer Security (TLS) and resource-constrained devices.
Paquin et al. \cite{paquin2020benchmarking} analysed the integration of quantum-resistant cryptography into TLS. 
Their benchmarking quantified computational and data transfer overheads, revealing the feasibility of PQC in web communication, though some algorithms increased overall network traffic due to larger key sizes and overheads. 
Notably, lattice-based techniques showed comparable results to classical methods, suggesting deployment of PQC is practical with the careful selection of algorithms (e.g., NIST standardisation efforts). 

Barton et al. \cite{Barton2019PostQC} also examined PQC integration in TLS, specifically focusing on constrained environments (similar to this study) and highlighting the inherent challenges. 
This pre-NIST standardisation work indicated potential additional PQC overheads (latency, computation, traffic) that vary with security level, compared to low-power classical cryptographic methods.

Tasopoulos et al. \cite{TLS13ResourceConstrained}, also pre-NIST, but evaluating many of its candidates, verified the resource utilisation of a complete TLS 1.3 implementation, and provided a valuable point of reference when validating the results of this paper. 
They found PQC could be implemented in resource-limited settings with lattice-based Kyber showing favourable performance. 
However, some other algorithms were less energy-efficient than the classical alternatives, emphasising the importance of good algorithm selection in implementation.


Schöffel et al. \cite{schoffel2021energy} specifically examined the energy costs of PQC Key Exchange Mechanisms (KEMs) in TLS-based low-power IoT devices. 
Their findings reinforced the significant energy cost variations among PQC techniques, reiterating the importance of selecting appropriate algorithms to minimise power consumption, based on both security needs and environmental constraints.


Finally, Roma et al. \cite{roma2021energy}'s analysis of PQC energy efficiency in mobile and large-scale environments showed performance variations based on vastly different architectures, but again highlighted the energy efficiency of lattice-based techniques. 
They concluded that algorithm selection should depend on specific requirements and platform characteristics, noting the increased significance of energy costs at scale and impact on battery life in portable applications. 


In summary, these studies collectively demonstrate the feasibility of integrating PQC in low-power devices, but consistently emphasise the need for careful algorithm and parameter selection.
This is particularly the case in resource-constrained environments to manage trade-offs in computation, memory, data transfer, and energy. 
Lattice-based techniques often appear efficient, aligning with early NIST selections. 
However, non-PQC algorithms may remain a valid choice when the implementation platform does not support post-quantum cryptography or where confidentiality or integrity are not deemed to be critical concerns and it may be de-prioritised based on risk. 

The following section details the methodology for researching the energy efficiency of key generation algorithms on the target hardware platform.




\section{Implementation} \label{Implementation}
NIST's PQC standardisation efforts are widely regarded as the most significant globally \cite{nist_pq_standards_FIPS203-205}, progressing through a series of competitive evaluation rounds \cite{nistir8547}. Submissions undergo rigorous peer review and are evaluated on their technical merits before selected algorithms progress to be finalised as standards.
These are the standards incorporated in this study.

\begin{figure}
  \centering
  \includegraphics[width=6cm]{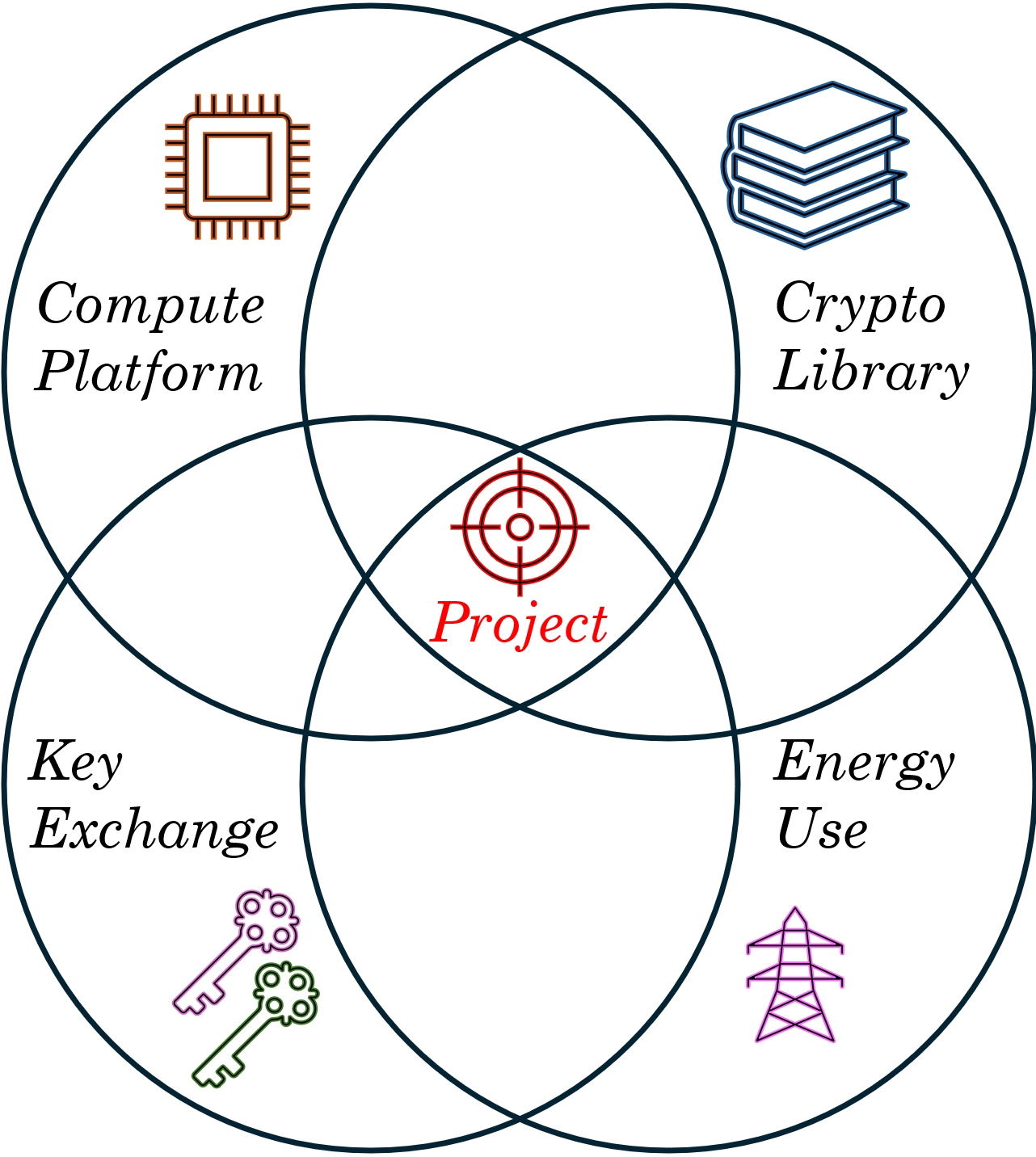} 
  \caption{Combination of four different subject areas: the Compute Platform (Pi), Library (OpenSSL3.5), Key Exchange (Classical+PQC ML-KEM), Energy Use (measured via a TC66C meter).}
  \label{4way}
\end{figure}

As indicated by the literature review, the field of post-quantum computing (PQC) is extensive. This paper focuses on a specific overlapping area defined by four key aspects:
\begin{enumerate}
  \item \textbf{Key Exchange:} A crucial component of secure communication whose traditional techniques were protected by large-number factorisation that will be rendered vulnerable to Shor's algorithm in quantum computing. New quantum-resistant techniques are emerging.
  \item \textbf{Energy Consumption:} A comparative evaluation of the energy used during key generation methods for both traditional and post-quantum cryptography.
  \item \textbf{Computing Platform:} The Raspberry Pi 5 has been selected as the test platform, ensuring homogeneity and repeatability of results. Its uniformly common integration into operational technology solutions and use by hobbyists makes it a relevant choice.
  \item \textbf{Software Library:} Standardised libraries are employed in solutions, recognising the implementation complexity and verification challenges of cryptographic code. Following established best practice in cryptography, this paper utilises the OpenSSL tested, peer-reviewed, and performant library rather than developing custom implementations. 
\end{enumerate}
An illustration of these four constituent parts is presented in Figure \ref{4way}.
This paper focuses on a practical comparison of the energy consumed during key generation for newly standardised PQC algorithms against traditional methods.
Experimentation is conducted on a Raspberry Pi platform, employing standardised cryptographic libraries to ensure a consistent basis for comparison. 
The aim is to establish a baseline for this platform to ascertain the impact of transitioning to PQC for this specific aspect of the cryptographic suite.

\subsection{Methodology and Implementation Overview}

To ensure the acquisition of reliable energy utilisation data, the experimental methodology involved executing key generation algorithms across a significant number of iterations for each algorithm.
Key generation was performed on the Raspberry Pi, with adjustments made for algorithms with longer key generation times to normalise each algorithm's time-on-test. 
Baseline energy consumption of the Pi platform (excluding key generation) was established through extensive NULL runs whose times were subsequently subtracted from experimental results including the key-generation stage. 

\begin{figure}
  \centering
  \includegraphics[width=1.0\textwidth]{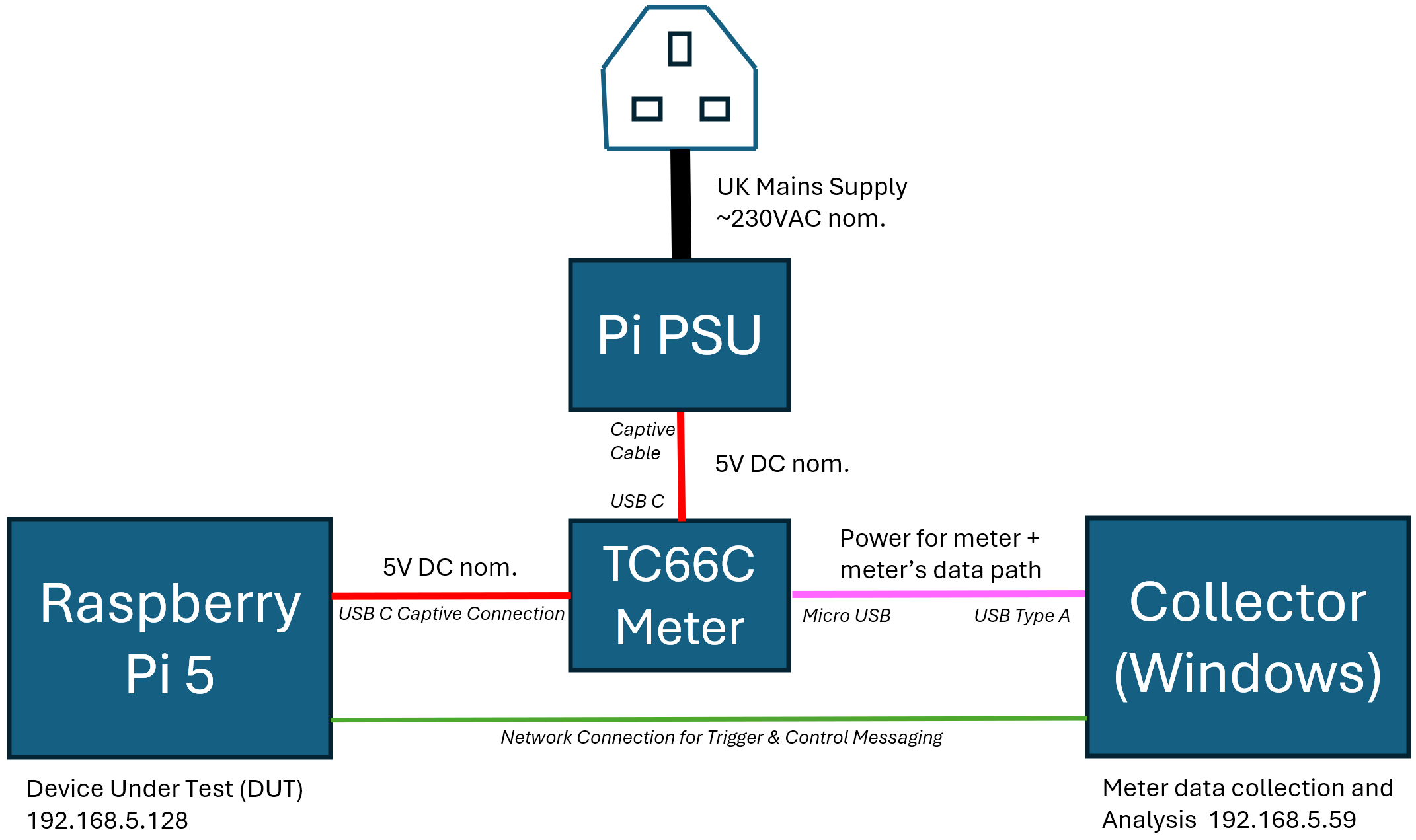}
  \caption{Block Diagram of the experimental setup - showing the Pi DUT which performs the key generation, and the Windows machine recording the electric supply characteristics from the TC66C meter. Trigger messages are conveyed using the network illustrated to START and STOP the measurements in line with the state of each experiment.}
  \label{SetupBlockDiagram}
\end{figure}

\begin{figure}[h!]
  \centering
  \includegraphics[width=6cm]{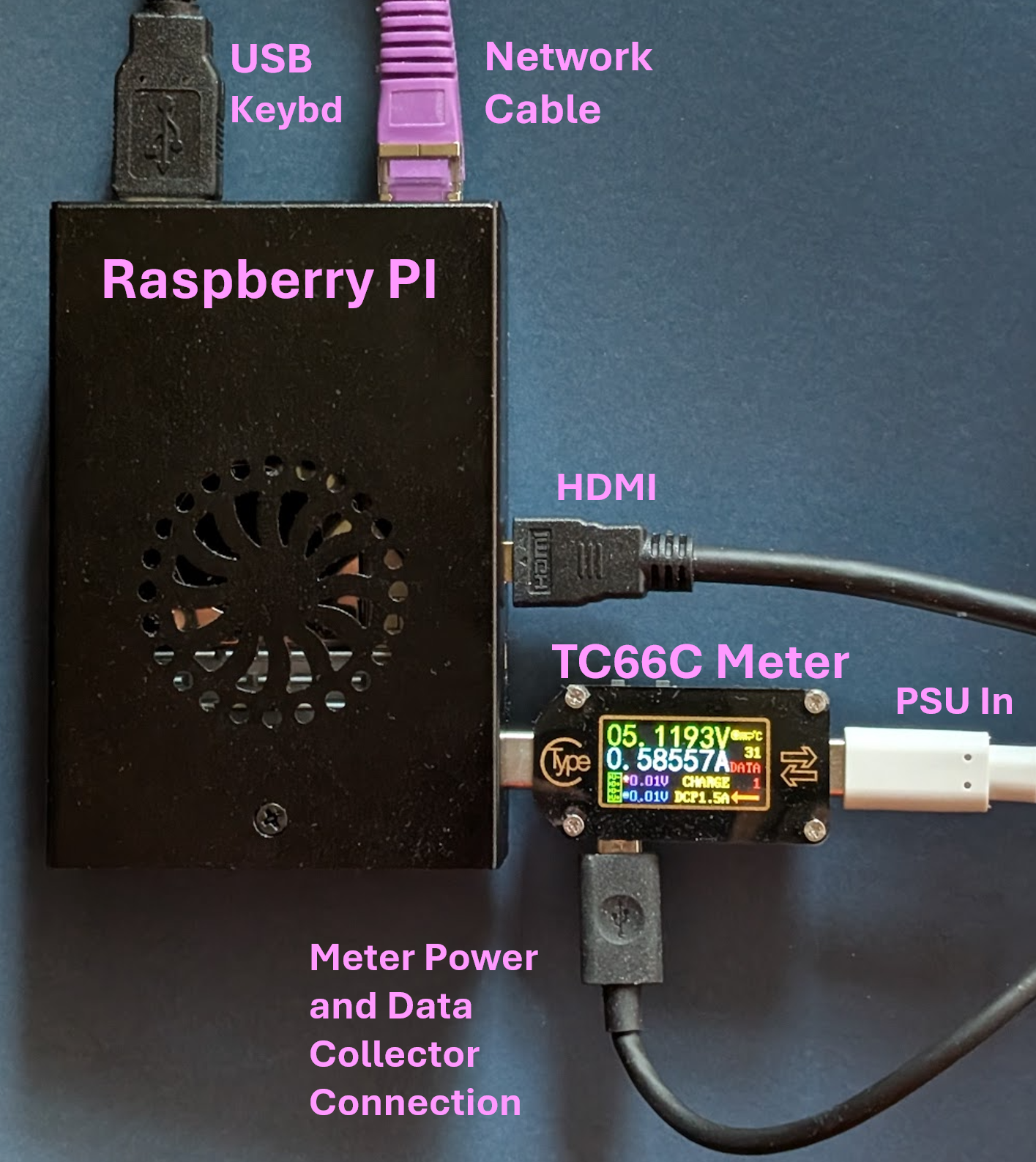}
  \caption{Photograph of the experimental setup illustrating the Raspberry Pi 5 Device Under Test (DUT) and the TC66C test meter inline with the Power Supply for the Pi, measuring its energy usage characteristics.}
  \label{pinkpi}
\end{figure}

A distributed architecture was employed for the experimentation, utilising a Raspberry Pi as the device under test (DUT) and a Windows-based data acquisition system to record electrical parameters from a TC66C energy meter inline with the Pi's power supply.
A block diagram of the experimental setup is illustrated in Figure \ref{SetupBlockDiagram}, and a marked-up photograph of the TC66C meter and Raspberry Pi in Figure \ref{pinkpi}.

Remote STOP/START signalling via network communication facilitated automated data collection on the Windows collector, with specific steps taken to maintain a consistent testing environment on the Pi, such as fixing the CPU clock and running the fan at 100\% as well as operating the testing over a very large number of iterations.
The software measurement developed for the Windows machine featured a multi-threaded design to manage network communication and data acquisition asynchronously to capture accurate results triggered by the signalling sent from the Pi DUT.



Aggregated full results are provided and analysed in Section \ref{Evaluation}, with practical examples of an experimental setup, including screenshots of what the user has visible to them in Section \ref{AppendixA} (Appendix A).
For reference, the source code, input and results data files are located on this paper's GitHub site \cite{CPGITHUB}.
Further details regarding the environment, experimental setup, system components, test procedures and software architecture are provided in Section \ref{Methodology}.

\subsection{Data Analysis}  \label{DataAnalysis}
Following each experiment the data collected in that algorithm's output file(s) are analysed to determine the DUT's energy consumption characteristics. 

\subsubsection{Calculating Energy Used}
The TC66C energy monitor returns a number of parameters across its USB Serial connection to the Windows collector.
Its power source is also from the Windows device avoiding influencing the passing power data recorded for the Pi.
The software created for this paper uses a TC66 software library provided by TheHWCave on GitHub \cite{HWCAVEGITHUB} to set up the connection to the meter and periodically recover highly accurate data on voltage, current, wattage and energy used.

Cumulative energy consumption is converted from milliwatt-hours (mWh) to Joules (J) thus:
\[
\text{Joules (J)} = \text{Milliwatt-hours (mWh)} \times 3.6
\]
The 3.6 factor is derived from there being 3,600 seconds in an hour, and 1,000 mWh in 1 Wh.


\subsubsection{Results Accuracy}

The TC66 unit used to gather the results in the experiments is not itself calibrated. 
However, an identical unit was tested against a calibrated Agilent lab supply in an analysis performed by TheHWCave \cite{thehwcaveyoutube}.
The results of this analysis found the tested TC66C unit to be well within its 0.05\% Voltage accuracy specification and the 0.1\% stated accuracy for current.
In the typical voltage and current window seen during experimental testing for this paper, TheHWCave's study demonstrated the meter exhibiting significantly better tolerances, typically within 0.02\% accuracy for voltage, and 0.05\% for current.


\subsection{Selection of Parameters Tested}

Our experimental methodology relied on several carefully chosen parameters, which we outline below along with their selection criteria:

\begin{itemize}
    \item \textbf{Iterations:} A range of experimental round sizes were performed in the environment.
    These were chosen to validate the results of the tests and the final outputs make use of the predominantly `500,000' key examples - with testing taking place over a period of around 24 hours in operation.  This iteration count is specified in the input file to the test script.

    \item \textbf{Security Levels:}  A range of algorithms have been chosen that cover the NIST security levels from 1 to 5, where a broad equivalence of the different algorithms under test have been made to permit easy comparison.

    \item \textbf{Algorithms:}  Three categories of algorithms are included in the testing, all are from the OpenSSL3.5 library as compiled for the Raspberry Pi device under test.  The full set can be seen listed in Table \ref{tab:security_levels}.
    \begin{enumerate}
        \item Elliptic Curve. Several commonly used ECC algorithms were chosen to enumerate across a range of security levels. This was used to baseline this method versus other key generation categories. 
        \item RSA. Another classic algorithm used for key generation. RSA is a more mature solution and still commonly in use. Over the years the key sizes used have ramped upwards to maintain security as computing power has risen over the years. Several key-sizes for RSA were selected to include in this analysis. 
        \item ML-KEM Post-Quantum Cryptography technique.  This is the currently available key generation technique standardised by NIST as FIPS 203, and implemented in the April 2025 OpenSSL3.5 library release. 
        \item The ML-DSA algorithm has also been included. While this is not used as a key exchange mechanism explicitly, it does perform key generation at equivalent security levels using Lattice techniques, and is therefore a useful yardstick to ensure the broad validity of results of the ML-KEM method, while awaiting the standardisation of HQC and its implementation in OpenSSL for evaluation.
    \end{enumerate}

    \item \textbf{Polling Interval:}
        This is a parameter which is chosen as part of the experimental setup.  This is the frequency at which data is recovered by the Windows collector machine from the TC66C meter, and does not impact on the key generation process. The data which is used for the energy calculations is a cumulative value from the TC66C, so the period chosen here defines only start and stop time granularity and the update period on the screen for the watching user.

\end{itemize}


\section{Methodology} \label{Methodology}

The experiments are designed to record empirical energy utilisation over a significant number of iterations to establish a reliable baseline. 
Key generation for each algorithm is executed multiple times across 10,000, 100,000, and 500,000 rounds - from under an hour to over a day in total duration. 
Where key generation times are considerably longer (for instance, with larger RSA key sizes), a lower iteration count is used, and the corresponding experiments are  adjusted accordingly too, balancing out time on test for each algorithm.

Furthermore, multiple NULL runs are performed in between key generation runs, involving identical numbers of loop iterations but triggering a 5ms delay instead of the OpenSSL key generation operation. 
The 5ms duration of the delay is not critical; rather, these runs serve to capture the baseline energy consumption of the Raspberry Pi platform (including the fan) incorporating everything other than the key generation itself. 
This background energy usage (averaged over multiple NULL runs) is then subtracted from the results obtained including the actual OpenSSL key generation processes.

Software has been developed to execute in the test environment, including cycling through the different key generation parameters, to trigger the start and stop of data capture from the energy meter and to provide feedback to the test operator on the status of the experiment.  
An example of live test runs and actual screen captures from them can be reviewed in Section \ref{AppendixA} (Appendix A).

Specific steps taken to ensure consistent measurement include running up the fan on the Pi to 100\% for all experiments, fixing the CPU clock to minimise the potential for dynamic frequency and voltage scaling influencing results, and pausing several seconds before starting the experiment to ensure the environment has settled particularly in relation to the fan speed. 

The test framework has been carefully designed to provide a consistent environment for each run and to ensure that accurate, repeatable experimental results can be gathered.  
The principle that the results gathering should not impact the device under test through viewing of its status is followed to develop the solution - minimising the opportunity to trigger `observer effect' such as through reporting using a separate collection machine rather than the device under test (DUT).

\subsection{Experimental Approach}
The experimental setup employs a distributed architecture to benchmark the energy consumption of a process running on the DUT (Raspberry Pi).
A Windows-based data acquisition system is utilised to record electrical parameters (voltage, current, power, etc.) from a Ruideng TC66 tester connected inline with the power supply of the Raspberry Pi.
The experiment is controlled remotely via network communication between the Raspberry Pi and the Windows machine, with trigger messages being sent for GETREADY, START and STOP purposes, aligning with the timing of the experiment running on the Pi.

The key system components include: \label{SystemComponents}

\begin{itemize}
    \item \textbf{Target System - Device Under Test - (Raspberry Pi):}
    This single-board computer will execute software which runs experiments whose energy consumption is being benchmarked.
    It will also host a network client responsible for sending control messages to accurately start and stop the energy acquisition recording.
    \item \textbf{Data Acquisition System (Windows Machine):}
    This system hosts the data logging software.
    It will be connected to the TC66C energy tester that monitors the electrical utilisation characteristics of the Raspberry Pi.
    This system will also run a network server to receive control messages to toggle acquisition on and off for the energy data.
    \item \textbf{USB Tester (Ruideng TC66C):}
    An external hardware device connected between the Raspberry Pi and its power supply.
    It will accurately measure and report instantaneous electrical parameters such as voltage, current, and power, and cumulative totals including energy consumed.
    \item \textbf{Ethernet Network:}
    The communication medium facilitating the exchange of control messages between the Raspberry Pi and the Windows machine. 
    This is a dedicated network with UDP messages used without the latency and additional energy of a TCP handshake.
\end{itemize}

\subsection{Experimental Procedure}  \label{ExperimentalProcedure}

A structured approach allows for remote control and automated data collection, enhancing the efficiency and reproducibility of the energy benchmarking experiments.
The experiment will proceed through the following stages, initiated and controlled via network communication (see also Section \ref{AppendixA} (Appendix A) for a practical example):
\begin{itemize}
    \item \textbf{Initialisation \& Preparation (Triggered by "GETREADY" message):}
        The Raspberry Pi, acting as the control client, will send a "GETREADY" message to the Windows data acquisition server.
        This message will include experimental parameters (e.g., test duration, sampling frequency, experiment identifier etc.).
        After receiving the "GETREADY" message, the Windows machine will:
        \begin{itemize}
            \item[-] Parse the experimental parameters.
            \item[-] Initialise the data logging system, preparing the output file(s) with appropriate headers based on the received parameters.
            \item[-] Transition the TC66C meter to "ready" state, awaiting the "START" command.
        \end{itemize}
    \item \textbf{Data Acquisition (Triggered by "START" message):}
        Once the process on the Raspberry Pi is ready for benchmarking (i.e. after the CPU clock is hard set and the fans have had time to spin up and settle), immediately prior to starting the experiment it will send a "START" message to the Windows machine.
        The Windows machine will then:
        \begin{itemize}
            \item[-] Initiate data collection from the USB tester at the specified sample rate.
            \item[-] Continuously record the measured electrical parameters (voltage, current, power, energy etc.) along with timestamps to the designated output file(s).
            \item[-] Simultaneously, a subset of the data is displayed on the Windows screen for real-time monitoring by the operating user (on the Windows machines, so not impacting the DUT).
        \end{itemize}
   \item \textbf{Termination (Triggered by "STOP" message):}
        At completion of the benchmarking process on the Raspberry Pi, it will send a "STOP" message to the Windows machine.
        The Windows machine will next:
        \begin{itemize}
            \item[-] Cease data acquisition from the USB tester.
            \item[-] Close the output file(s), ensuring all data is saved.
            \item[-] Cleanly terminate the data logging application and network server thread.
        \end{itemize}
    \end{itemize}

\subsection{Software Design / Architecture}  \label{SoftwareArchitecture}
The software on the Windows machine is designed with a multi-threaded architecture to ensure responsiveness and separation of its asynchronous roles:
\begin{itemize}
    \item \textbf{Main Control Thread:}  \label{MainControlThread}
    This thread will be responsible for coordinating the overall operation of the data acquisition.
    It will also control the other threads and overall flow.
    \item \textbf{Network Server Thread:}  \label{NetworkServerThread}
    This will continuously listen for incoming network messages ("GETREADY", "START", "STOP") from the Raspberry Pi, and will be started and stopped by the main thread.
    After receiving a message, it will parse the command and any associated parameters, signalling the main control thread to take appropriate action.
    \item \textbf{Data Acquisition Thread:} \label{DataAcquisitionThread}
    This thread will be responsible for periodically reading data from the USB tester.
    It will also write the data to the output file(s) and display pertinent information on the screen to give positive feedback to the user/operator as to current operational state.
    This thread will be started and stopped by the main thread based on the "START" and "STOP" messages received via the network from the Pi.
\end{itemize}

\section{Evaluation} \label{Evaluation}

Table \ref{tab:security_levels} presents a categorisation of the key generation algorithms employed by this study, mapped to their equivalent NIST Security Levels interpretted by the author from NIST SP 800-57 Part 1 Revision 5 \cite{nistsp800-57p1r5}. 
The selection of algorithms provides coverage across all five NIST security level equivalents, encompassing established classic cryptography with RSA and Elliptic Curve Cryptography (ECC), and the emerging post-quantum cryptography standards.
Module-Lattice-Based Key-Encapsulation Mechanism (ML-KEM) as specified in FIPS 203 (Kyber) is the primary algorithm on test with the key generation part of ML-DSA for check-and-balance purposes. 

For clarity, common alternative names and relevant FIPS standards are included in brackets alongside the protocol identifier where applicable. 
Additionally, Table \ref{tab:security_levels} provides the estimated equivalent symmetric key size in bits for each algorithm, offering a comparative measure of their current perceived security strength.

\begin{table}[h!]
\centering
\caption{NIST Security Levels and Equivalent Symmetric Key Sizes for OpenSSL Key Generation Protocols (Based on NIST SP 800-57 Part 1 Rev. 5 (Table 2) \cite{nistsp800-57p1r5} and the author's categorisation).}
\label{tab:security_levels}
\begin{tabular}{@{}p{1.4cm}p{5cm}p{3cm}p{1.3cm}<{\raggedleft}@{}}
\toprule
\textbf{NIST Sec Level}  & \textbf{Protocol}     & \textbf{Category}    & \textbf{Equiv. Bit Size} \\
\midrule
Level 1             & RSA-1024              & Classic Technology         & 80                                   \\
\midrule
Level 2             & EC secp160r1          & Elliptic Curve Cryptography & 80                                   \\
                    & RSA-1536              & Classic Technology         & $\sim$96                             \\
\midrule
Level 3             & EC secp224r1          & Elliptic Curve Cryptography & 112                                  \\
                    & RSA-2048              & Classic Technology         & 112                                  \\
                    & EC P-256 (secp256r1, FIPS 186-4) & Elliptic Curve Cryptography & 128                                  \\
                    & ML-DSA-44             & Post-quantum Technology    & $\sim$128                            \\
                    & ML-KEM-512 (Kyber-512, FIPS 203) & Post-quantum Technology    & $\sim$128                            \\
\midrule
Level 4             & RSA-3072              & Classic Technology         & 128                                  \\
                    & EC P-384 (secp384r1, FIPS 186-4) & Elliptic Curve Cryptography & 192                                  \\
                    & ML-DSA-65             & Post-quantum Technology    & $\sim$192                            \\
                    & ML-KEM-768 (Kyber-768, FIPS 203) & Post-quantum Technology    & $\sim$192                            \\
\midrule
Level 5             & RSA-4096              & Classic Technology         & $\sim$140                                  \\
                    & EC P-521 (secp521r1, FIPS 186-4) & Elliptic Curve Cryptography & 256                                  \\
                    & ML-DSA-87             & Post-quantum Technology    & $\sim$256                            \\
                    & ML-KEM-1024 (Kyber-1024, FIPS 203) & Post-quantum Technology    & $\sim$256                            \\
\bottomrule
\end{tabular}
\end{table}

\subsection{Energy Results - Key Generation}
The experiments were conducted on a Raspberry Pi 5 platform under previously outlined controlled conditions, with fixed clock speeds and 100\% fan to establish a consistent baseline for energy consumption measurements across each of the selected algorithms.

\begin{figure}[h!]
  \centering
  \includegraphics[width=12cm]{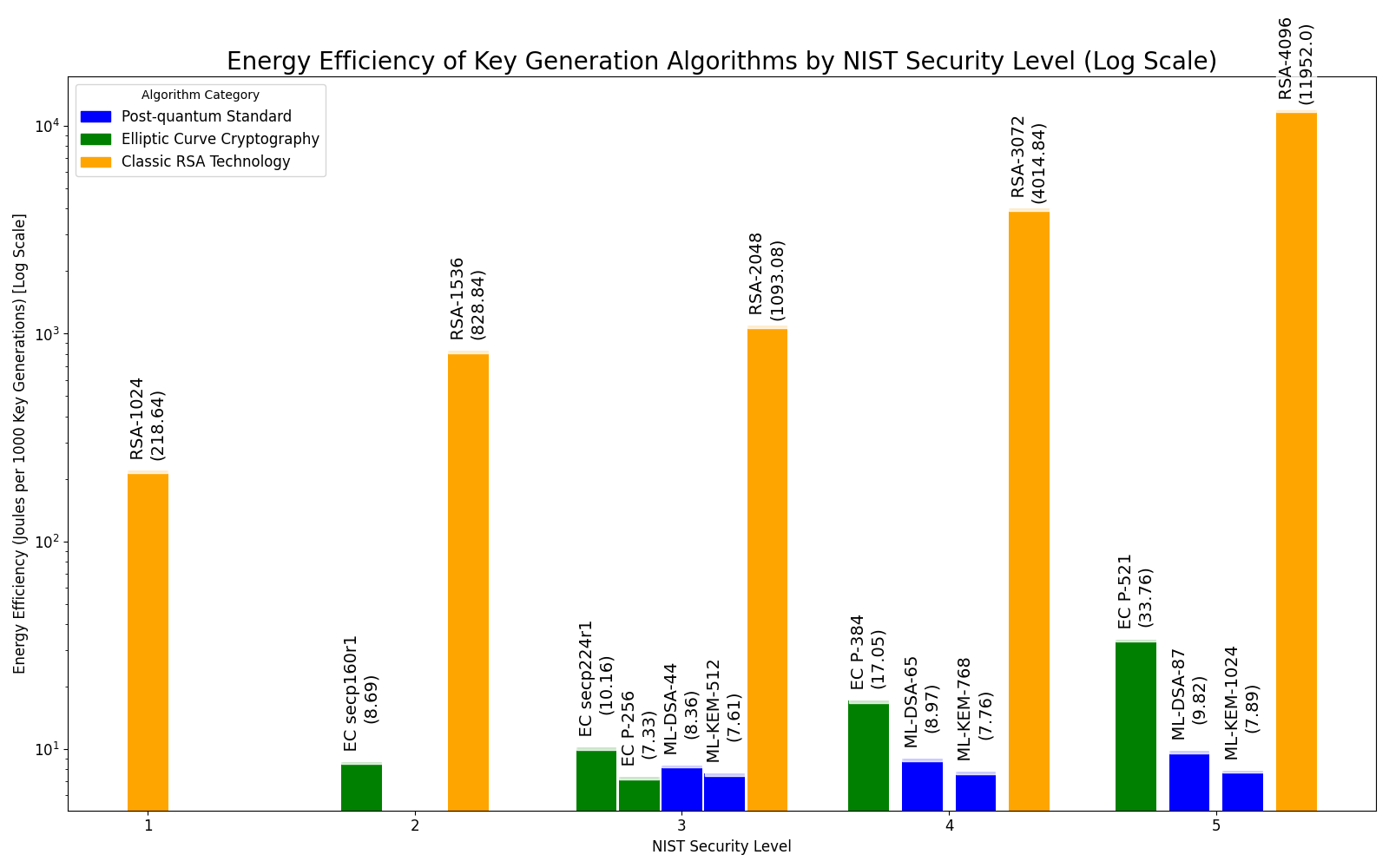}
  \caption{Plot of Energy Rate in Joules/1,000 key generations of each algorithm, categorised by NIST security level and type of algorithm. Experiment performed over a typical 500,000 key generation count for each algorithm.  (A tablulated version is provided by Table \ref{tab:energy_levels}).}
  \label{EnergyComparePlot}
\end{figure}

The bar chart in Figure \ref{EnergyComparePlot} visualises the energy efficiency of several key generation algorithms across different NIST Security Levels.
The horizontal axis represents the NIST Security Level, ranging from 1 to 5. 
There may be one or more algorithm in each level.
The vertical axis displays the energy efficiency in Joules per 1,000 key generations, using a logarithmic scale to accommodate the wide range of values, particularly relating to RSA.
Each bar on the chart corresponds to a specific key generation algorithm.
The colour of each bar indicates the category of that algorithm: blue represents the recently standardised post-quantum algorithms, green:  elliptic curve cryptography, and orange: classic RSA technology.
The name of each algorithm and its efficiency is printed above the corresponding bar for identification and comparative purposes. 

\begin{table}[h!]
\centering
\begin{tabular}{@{}p{1.4cm}p{5cm}p{3cm}p{1.3cm}<{\raggedleft}@{}}

\toprule
\textbf{NIST `Sec' Level}  & \textbf{Protocol}     & \textbf{Category}    & \textbf{J/1,000 keygens} \\
\midrule
Level 1             & RSA-1024              & Classic Technology         & 218.64                                   \\
\midrule
Level 2             & EC secp160r1          & Elliptic Curve Cryptography & 8.69                                   \\
                    & RSA-1536              & Classic Technology         & 828.84                             \\
\midrule
Level 3             & EC secp224r1          & Elliptic Curve Cryptography & 10.16                                  \\
                    & EC P-256 (secp256r1, FIPS 186-4) & Elliptic Curve Cryptography & 7.33                                  \\
                    & ML-DSA-44             & Post-quantum Technology    & 8.36                            \\
                    & ML-KEM-512 (Kyber-512, FIPS 203) & Post-quantum Technology    & 7.61                            \\
                    & RSA-2048              & Classic Technology         & 1093.08                                  \\
\midrule
Level 4             & EC P-384 (secp384r1, FIPS 186-4) & Elliptic Curve Cryptography & 17.05                                  \\
                    & ML-DSA-65             & Post-quantum Technology    & 8.97                            \\
                    & ML-KEM-768 (Kyber-768, FIPS 203) & Post-quantum Technology    & 7.76                            \\
                    & RSA-3072              & Classic Technology         & 4014.84                                  \\
\midrule
Level 5             & EC P-521 (secp521r1, FIPS 186-4) & Elliptic Curve Cryptography & 33.76                                  \\
                    & ML-DSA-87             & Post-quantum Technology    & 9.82                            \\
                    & ML-KEM-1024 (Kyber-1024, FIPS 203) & Post-quantum Technology   & 7.89                            \\
                    & RSA-4096              & Classic Technology         & 11952.0                                  \\
\bottomrule
\end{tabular}
\caption{Equivalent to Figure \ref{EnergyComparePlot} with numerics to compare instead of logarithmic bar heights.  RSA is particularly resource-hungry at larger key-sizes.}
\label{tab:energy_levels}
\end{table}

\subsubsection{Energy Results - Commentary}
The Post-Quantum Cryptography techniques are broadly equivalent in terms of their energy utilisation when compared with the Elliptic Curve Cryptography methods at NIST Security Level 3.  
At NIST levels 4 and 5 then the post-quantum techniques are more energy efficient for the same security level than the tested current ECC methods.
Turning to RSA, as expected the energy efficiency becomes exponentially worse as the key size increases.  
Compared with ECC, RSA is around two or more orders of magnitude less efficient at key generation as the elliptic curve equivalent, plus generates larger keys.
Comparing RSA to the PQC techniques shows an even greater offset, as the PQC is much flatter across the resources it requires as the security level increases.



\subsubsection{Extrapolating Potential Energy Savings}

This section presents a highly speculative, order-of-magnitude estimate of the potential energy savings that could be realised by transitioning notionally from RSA-2048 key generation to ML-KEM-512 on a large scale. 
Due to the decentralised nature of key generation across various systems globally, the assumptions made here are necessarily quite broad.
Key generation too is a comparatively low-volume cryptographic operation compared with others in the suite, and the results are accordingly small.                                      

Several key areas where RSA key generation is prevalent are considered:

\begin{itemize}
    \item \textbf{Web Servers (HTTPS):} Assuming approximately 300 million active websites globally using certificates (an estimate in Jan 2025 based on \cite{ssldragon2025sslstats}), and a conservative average certificate cycle of 3 months (LetsEncrypt uses 60 days and has over 50\% of the market), this equates to roughly 1.2 billion key generations per year.
    \item \textbf{VPN Users:} With an estimated 1.5 billion VPN users globally \cite{vpnstats2025}, assuming each host generates a new key each year, this adds another 1.5 billion key generations annually.
    \item \textbf{Other Applications (VPN, Email, SSH):} A conservative estimate for the combined key generation for secure email, SSH, and other applications is around 10\% of the web server figure, supposing approximately 120 million key generations per year.
\end{itemize}

Based on these highly speculative assumptions, the total estimated RSA key generations per year could be in the order of 2.82 billion.

The experiment demonstrates (Figure \ref{EnergyComparePlot}) that the energy consumption for generating a single RSA-2048 key is approximately 1.093 Joules on the test platform. 
Therefore, taking this and RSA-2048 as the baseline, the total estimated annual energy consumption for key generation across these applications if performed on a Raspberry Pi 5 would be in the order of 3.082 GigaJoules.

To convert this to kilowatt-hours (kWh):
\[ 3,082,000,000 \, \text{J} \times \frac{1 \, \text{W} \cdot \text{s}}{1 \, \text{J}} \times \frac{1 \, \text{W} \cdot \text{h}}{3600 \, \text{s}} \times \frac{1 \, \text{kW} \cdot \text{h}}{1000 \, \text{W} \cdot \text{h}} \approx 856.11 \, \text{kWh} \]

Assuming an average electricity unit price of £0.26 per kWh (based on the UK average from QEP Table 2.2.4 referenced in: \cite{govukenergyprices}), the estimated annual cost for RSA-2048 key generation across these 2.82 billion keys would be approximately:
\[ 856.11 \, \text{kWh} \times 0.26/\text{kWh} \approx 222.59  GBP\]

The experiment's determined energy consumption for ML-KEM-512 (offering comparable NIST Level 3 security to RSA-2028) to be approximately 7.61 Joules per 1,000 key generations, or 0.00761 Joules per key. The energy saving per key generated would therefore be $1.093 - 0.00761 = 1.09939$ Joules for each key. This suggests that less than 1\% of the energy would be required for key generation using ML-KEM-512 PQC compared with RSA-2048, $\sim$131 times more efficient. 

For NIST `Level 5' security, comparing RSA-4096 (11.952 Joules per key) with ML-KEM-1024 (approximately 0.00789 Joules per key), the potential energy saving multiplier is even more significant, at around 1,500 times, whereas ECC P-521 (approximately 0.003376 Joules per key) is around 350 times more energy efficient than RSA-4096. Noted also here is that ML-KEM-1024 is circa 4 times more efficient than EC-P521, a significant advantage for the newer quantum-robust technology.


In conclusion, while this extrapolation is based on numerous broad assumptions, it suggests that a large-scale transition from traditional key generation techniques like RSA-2048 to post-quantum alternatives such as ML-KEM could lead to substantial energy savings and reduced computational overhead as well as protection from quantum techniques such as Shor. Additional research with more granular usage statistics is required to further refine these estimates.


\subsection{Key Generation Time Observation}

\begin{figure}[h!]
  \centering
  \includegraphics[width=12cm]{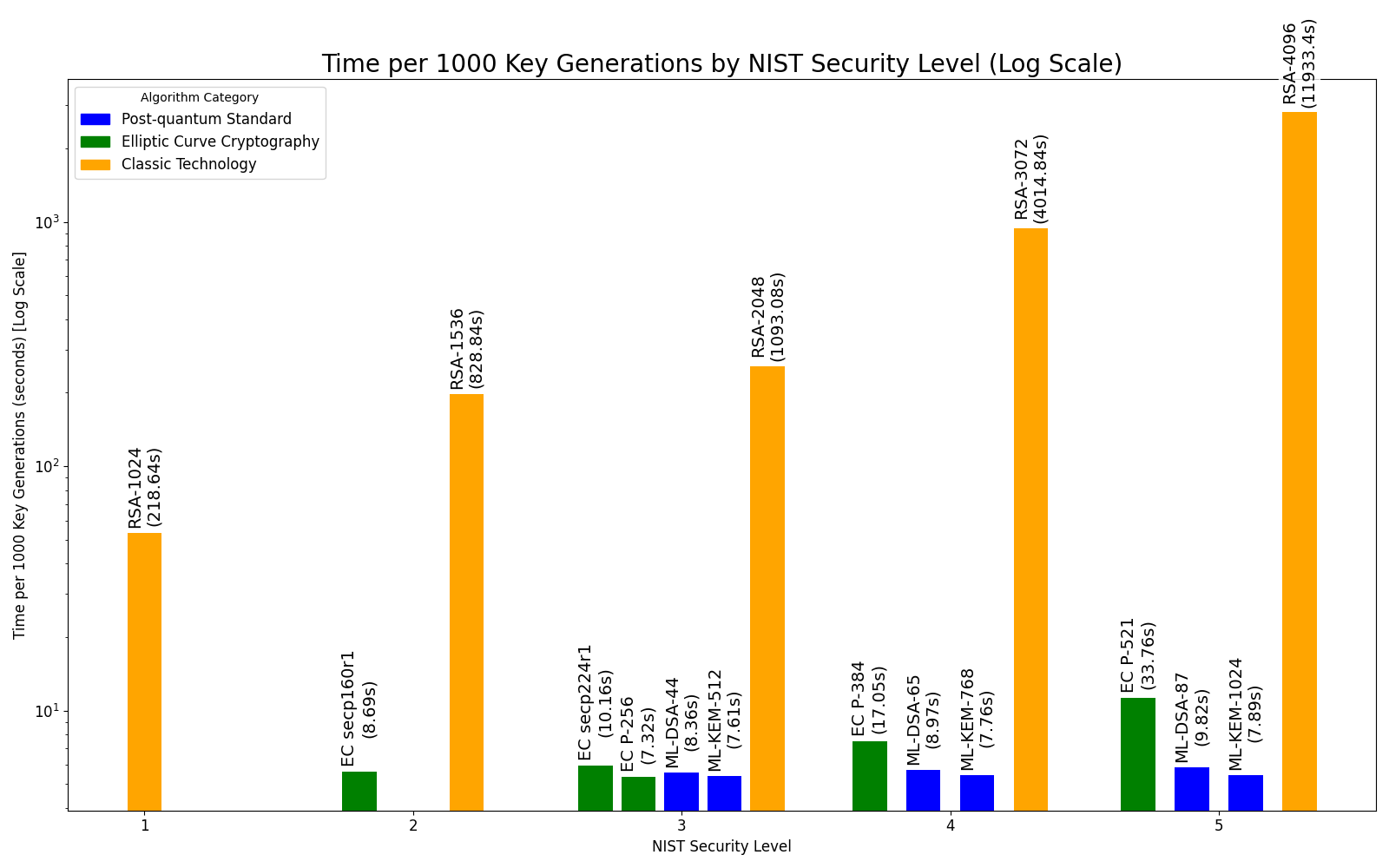}
  \caption{Similar plot to Energy Utilisation in Figure \ref{EnergyComparePlot}, but detailing time to generate 1,000 keys for each algorithm on the Pi 5 Device Under Test.}
  \label{TimeComparePlot}
\end{figure}

Whilst not the goal of this paper, the plot of compute time can additionally be created and studied, purely for academic interest. This is illustrated in Figure \ref{TimeComparePlot}.
The compute time required per 1,000 key generations exhibits a broadly similar shape to the energy consumption graph (Figure \ref{EnergyComparePlot}), with RSA algorithms generally requiring significantly longer processing times, particularly at higher NIST Security Levels.
PQC again performs well against the ECC category methods, outperforming them as the security level rises in comparison.

However, the relative performance in compute time between RSA and the two other categories differs. 
It appears that RSA consumes energy at a slower rate than for the ECC and PQC techniques, leaving a larger time multiple for RSA versus the other techniques.

A tentative conclusion is that RSA may not be fully optimised in OpenSSL3.5 on the Pi, or that it is less able to be parallelised with more serialisation portions exhibiting Amdahl's Law \cite{Amdahl1967}.
This finding is included as an observation, but is not explored further in this paper as it is not the main purpose of the study.
However the author believes it to be an interesting observation, still worthy of note.

\section{Conclusions} \label{Conclusions}

This paper's experimental analysis strongly suggests that a shift towards post-quantum key generation cryptography is desirable for new deployments due to its comparable or superior energy efficiency and inherent resilience to future quantum threats.
In Figure \ref{EnergyComparePlot} it is clear that as the NIST security level increases, the lattice-based key generation methods outperform ECC, and significantly outperform RSA, whilst adding resilience to known and projected post-quantum computing attacks.

\subsection{RSA's dead-end}
It is particularly apparent that RSA's viability is increasingly challenged, even before the potential impact of quantum computing on its integrity is taken into account.
Key sizes for RSA already need to be 2048-4096 bits and beyond for best practice (NIST Level 3 to Level 5), due to the impact and roadmap of classical computing progression (Moore's Law \cite{MooreSchaller}), and the cryptographic attacks this enables. 

The energy used for RSA-4096 key generation is orders of magnitude greater than for NIST security level 5 `equivalents' such as Elliptic Curve P521 and ML-KEM-1024 (see Figure \ref{EnergyComparePlot}).
When the potential of quantum computing algorithms is taken into account, even the Level 3 - Level 5 key sizes for RSA do not provide sufficient protection.
Larger RSA key sizes (towards 1TB) that do offer additional resilience to PQC attacks \cite{1TBRSA} quickly become unsustainable in terms of their requirements for memory and consumption of computational resource and energy.

\subsection{Use of Elliptic Curve Cryptography}
From the perspective of Elliptic Curve Algorithms at NIST Level 3, the Post-quantum ML-KEM methods remain close to the energy used by the classical ECC method (see Figure \ref{EnergyComparePlot} and Table \ref{tab:energy_levels}).
However, ML-KEM has a discernible advantage at NIST security Levels 4 and above in the tested environment.
When comparing with RSA, ECC should be preferred where it is still necessary to use a classical technique from a purely energy-based perspective where this option is available.
Use of such algorithms is likely to be prevalent for some time, especially as the NIST post-quantum standards \cite{nist_pqc_round4_status} have only recently been published (2024-2025) and are not yet commonly deployed or available in mainstream hardware and software libraries \cite{openssl35}.

\subsection{Adoption of New PQC Standards}
Considerable lead times will still be required for the implementation and testing of libraries based on the new PQC standards for deployments.
Integration into the software stacks of general-purpose computing is only just beginning (as evidenced by the April 2025 v3.5 production release of the widely-used OpenSSL library, which incorporates PQC methods \cite{openssl35}).
It will also take some time to incorporate any methods that can be hardware-accelerated onto general-purpose and application-specific processors, now that some standards have been ratified.

The PQC standards are also not static. They are still being developed, with new rounds of evaluation being led by NIST to diversify approaches beyond lattice techniques \cite{nist_pqc_round4_status}.
Even once the full set of FIPS standards are finalised, there will remain a long tail of platforms to be updated, particularly in the operational technology and embedded space which have extended deployment lifetimes, prioritising availability over other most else \cite{CentreonOTUptime}.

Governmental guidance is also relatively new in this area, outlining recommended roadmaps to quantum-safe cryptography within their jurisdictions \cite{cyber_gov_au_cryptography,ncsc_pq_timelines,nistir8547,eu_pqc_joint_statement_2024}.
It too will take some time to incorporate PQC into practice and into regulatory requirements for specific industry sectors.

\subsection{Recommendations}
This paper has experimentally analysed the energy characteristics of classical versus PQC key generation and evaluated the results obtained in this area.
The author recommends that classical key generation techniques should no longer be considered for new implementations, as the energy efficiency results are clear.
Non quantum-safe methods should be phased out of existing applications and deployments as is feasible, especially given that the post-quantum alternatives have been shown to match or exceed energy efficiency performance in the tested environment – by significant margins in many instances.

\subsection{Future Work}
The recommendation above is qualified by further potential experimental work utilising this test platform and software to broaden its scope. This would enable a more comprehensive end-to-end study to be performed with more parameters and to understand the full impact of the transition to PQC compared to classical cryptographic techniques.
This scope may be expanded in a number of suggested ways:
\begin{itemize}
    \item \textbf{Additional Hardware Platforms:} To use and compare more than one type of hardware, for example adding Raspberry Pi 4 and another standard Intel x86/AMD64 ISA platform for comparisons. 
    The meter used (unmodified) can capture up to 20 V with currents up to 5 A. Devices with attached batteries (e.g., laptops) are not suitable for this exercise due to the need to capture live energy utilisation directly.  
    \item \textbf{Alternative Software Libraries:} To identify further suitable cryptographic libraries for comparative testing, other than OpenSSL.
    For example to explore Bouncy Castle \cite{bouncycastle} and/or other potential PQC libraries as they become available. 
    A challenge here will be ensuring comparable algorithm support and maturity across different libraries. 
    Possible alternative sources may be found directly at asecuritysite \cite{asecuritysite_55130} which is a cryptographic website with resources for researchers, hobbyists and professionals in the area, including for Post-Quantum Cryptography.
    \item \textbf{Explore End-to-End Sessions:} 
    To expand the study to a fuller E2E lifecycle measurement across a range of scenarios:
    \begin{enumerate}
        \item On-Network Key Exchange:  Measure the bidirectional energy consumption of different PQC and
        traditional key exchange protocols including network communication.
        \item Data Transfer:  Quantify the energy cost of encrypting and transmitting
        datasets of varying sizes using the negotiated symmetric keys derived from both PQC
        and traditional key exchange mechanisms (also considering the cost of the required hardening to AES-256 for all applications that Grover's algorithm effectively mandates).
        \item Digital Signatures: Evaluate the energy efficiency of generating and verifying
        digital signatures using standardised PQC algorithms (e.g., ML-DSA, SLH-DSA)
        comparing to traditional signature schemes. Isolating the energy cost of specific 
        operations in an end-to-end scenario will be a key challenge to this potential future work.
    \end{enumerate} 
    \item \textbf{New Standards:} To incorporate the evolution of the standards, for example to include HQC testing once this has been finalised by NIST \cite{nist_pqc_round4_status} into a FIPS standard, and is incorporated into OpenSSL and other cryptographic libraries. Additional work to evaluate this algorithm could be carried out upon its availability, even whilst in Beta testing.
\end{itemize}

\begin{figure}[h!]
  \centering
  \includegraphics[width=9cm]{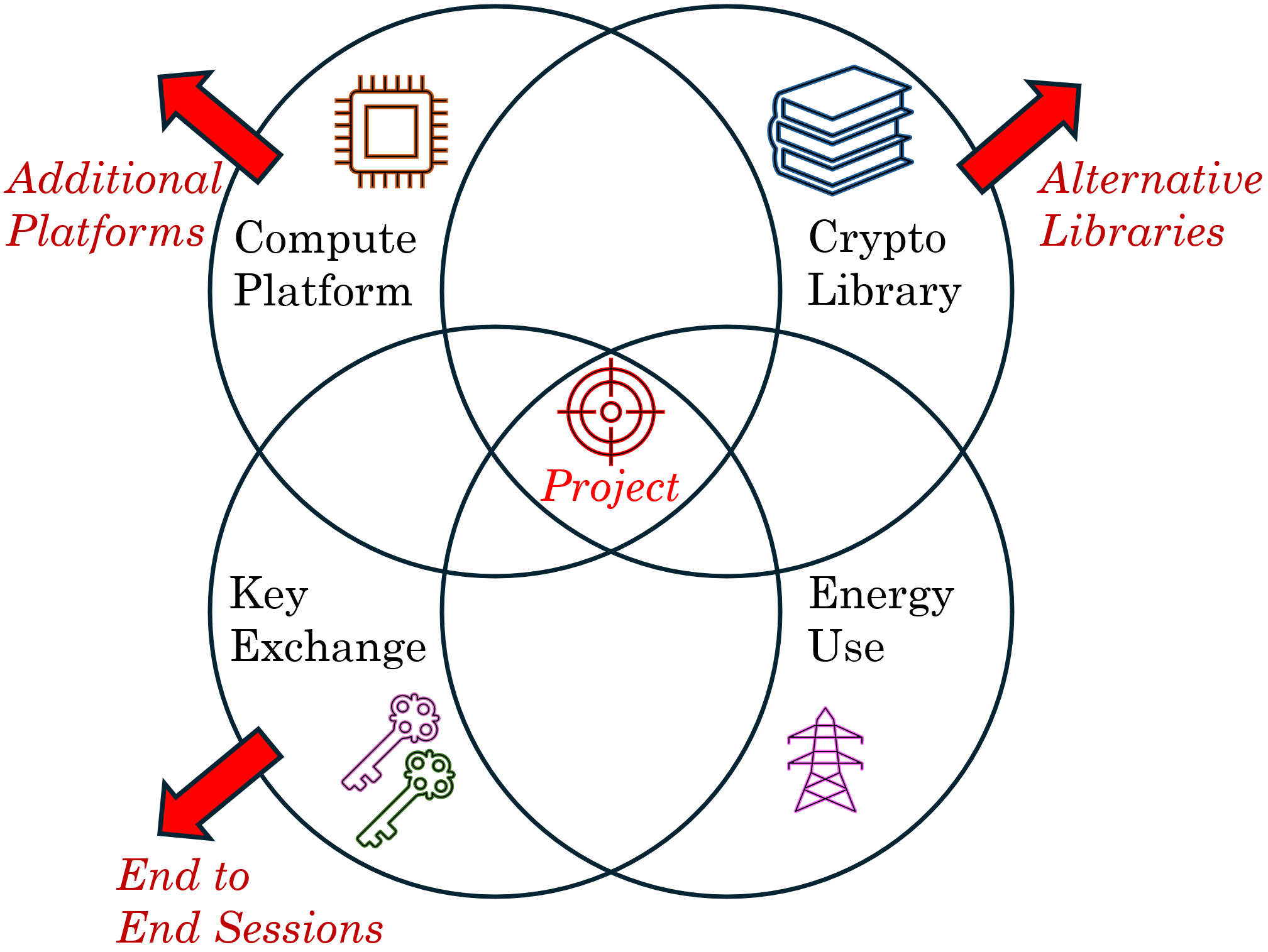}
  \caption{Future work scope expansions possibilities for this research project.}
  \label{FutureVenn}
\end{figure}

The experimental broadening proposed by Future Work and how it impacts on the areas selected for the initial scope of this study is illustrated in Figure \ref{FutureVenn}.

A significant portion of the work for this paper involved investing time in developing a robust methodology for capturing the energy utilisation data. 
The setup is outlined in more detail in Section \ref{SoftwareArchitecture}. 
This framework has been designed to allow this study to be extended in the future to other applications that utilise similar experimental setups. 
Having established energy efficiency differences in key generation, future research should examine if the patterns of results persist across all operations and end-to-end encrypted communication protocols.













\section{Appendix A - Practical Setup and Operation} \label{AppendixA}

This section has details of the equipment provisioned as part of the framework, and examples of the process and screens during a practical capture session.  
It is provided to ensure that the setup can be replicated and results verified as necessary. 

\subsection{Test Equipment Details}
Referencing the block diagram of Figure \ref{SetupBlockDiagram}, there are three major components in the setup:
\begin{itemize}
    \item The Raspberry Pi which is performing the key generation experiments
    \item The Ruideng TC66C energy meter, inline with the Pi's PSU, measuring supply parameters
    \item The Windows Collector Machine which is powering the TC66C meter and recording the DUT's energy use data between the START and STOP command messages.
\end{itemize}
  
\noindent Further details of the configuration and setup of these components follows.

\subsubsection{Raspberry Pi}
To allow for its replication, the Raspberry Pi has the following configuration of note:

\begin{itemize}
    \item Raspberry Pi 5 Model B Rev 1.0 8GB.
    \item Raspberry Pi OS Lite (Fully Patched as of 1 April 2025).
    \item OS `PRETTY-NAME'="Debian GNU/Linux 12 (bookworm)".
    \item Added software:  OpenSSL 3.5 and dependencies.
    \begin{itemize}
        \item Download, unpack, local compilation is required. Add OpenSSL3.5 binary to the path for system. Switch default openssl version to v3.5.
    \end{itemize}        
    \item The `batch\_experimenter.py' \& `experimenter.py' files and example source files can be obtained from the project GitHub page \cite{CPGITHUB}.
    \item This software sets the Pi fan to 100\% and fixes the clock frequency at the maximum to ensure these are as invariant as possible during the experiments and iterated tens of thousands of times.
\end{itemize}

\noindent Consistently attach to the Pi via SSH or via standard keyboard, mouse and video connections to run the experiments. 
In the experimentation for this paper SSH was used.
See Figures \ref{SetupBlockDiagram} and \ref{pinkpi} for how other connections are made. 
On the network port the IP address should be manually set, as it is directly connected to the Windows collector machine and no DHCP server is available. 
Example IPs are embedded in the code and on the block diagram (Figure \ref{SetupBlockDiagram}).

\subsubsection{Ruideng TC66C Meter}
Connect the meter as indicated in Figures \ref{SetupBlockDiagram} and \ref{pinkpi} . The configuration is as follows:

\begin{itemize}
    \item Ruideng TC66C Meter.
    \begin{itemize}
        \item It can also be the non-C version as Bluetooth is not required. BT can be turned off in the interface. The data is collected by USB on the Collection PC.
    \end{itemize}
    \item Device Switches: Power and PD toggle switches should both be OFF.
    \begin{itemize}
        \item Power switch = off, indicates that the unit should be powered from the Collection PC USB connection, and not from the Pi power source. 
        \item PD switch = off, as there should be no negotiation by the meter to the Pi Power supply - this data should pass-through natively.
    \end{itemize}
    \item The meter is using the v1.14 firmware that it was delivered with.
\end{itemize}

\noindent Use a Micro USB cable to attach the collection PC to the meter. 
This powers the device too, so its consumption is not part of the recorded data and it provides a virtual COM serial connection. 
The project's GitHub software running on the Connection PC will try to auto-detect the COM ports available, and the user will have the opportunity to check and select the correct option or, if known in advance, specify it via a command line switch bypassing the selection process for expediency.

\subsubsection{Windows Data Collection PC}
Connect as per Figure \ref{SetupBlockDiagram}, with the USB connecting to the MicroUSB on the meter, and an Ethernet connection to the Pi using a static IP address. 
In the experimental setup for this paper, the desktop PC used had a high specification - this is absolutely not required. 
The responsibilities of this machine are in logging data from a USB serial connection, so are relatively lightweight.  
It's pertinent hardware and software configuration is as follows:
\begin{itemize}
    \item Windows PC - the experiment's PC was running Windows 11 Professional 24H2, Build: 26100.3775. Fully Patched through 7th May 2025.
    \item A USB connection able to attach to the TC66C meter, using a generic Microsoft Serial USB Adaptor device driver.
    \item Python 3.x - the test PC used Python v3.13.3.
    \item Pip packages for `pycryptodome' and `pyserial' to communicate to the TC66C meter.
    \item Running the `responder.py' code from the GitHub project repository. 
    \item Trivial PC hardware needs, but for transparency the test Collection PC was an AMD Ryzen 9 7900X 12C 24T Processor with 96GB DDR5 RAM and 2TB SSD.
    \item Size the Collection PC according to the requirements:
    \begin{itemize}
        \item The PC is running a master control thread which spawns two additional threads for networking and polling data from the TC66C meter.
        \item Listening and responding to Network messages for START and STOP commands. Reading a small segment of data from the USB/Serial connection periodically at less than or equal to 10Hz. Printing a short update to the screen to keep the user updated on status. Writing out single lines of data to files at a maximum frequency of 10Hz.
    \end{itemize}
\end{itemize}

It would be relatively trivial to convert the Collection software to permit a Linux based collection device (or Apple) if required.  
The current test to ensure the device is a Windows PC could be adjusted to adapt to serial port identifications and file permissions on other platforms without requiring platform-specific versions. 

During experimentation, attach to the collection PC via SSH or via standard keyboard, mouse and video connections to run the required scripts and see the energy collection outputs from the experiments.  
An example of interaction using this framework is provided in the following section for reference.

\subsection {Example Experimental Run on the Framework}
In this section both the inputs and outputs from an experiment on the test platform are included.  
Firstly, a 3,200 iteration key generation test is carried out with algorithm ML-KEM-1024, demonstrating the expected output from the Pi and PC collector platforms.  
Secondly, a short batch algorithm test is carried out, again demonstrating expected outputs from the system.

\subsubsection{Single Run}
\begin{figure}[h!]
  \centering
  \includegraphics[width=12cm]{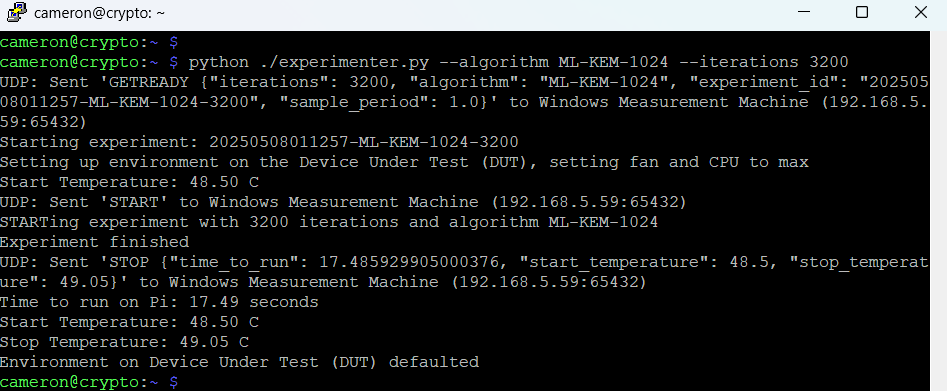}
  \caption{Command Line from the Pi Experimenter platform - 3,200 iteration ML-KEM-1024.}
  \label{PiSingle}
\end{figure}

In Figure \ref{PiSingle} the command `python ./experimenter.py \text{-}\text{-}algorithm ML-KEM-1024 \text{-}\text{-}iterations 3200' executes 3,200 key generations of ML-KEM-1024, a NIST level 5 algorithm.  
A screen capture is provided as Figure \ref{PiSingle}, and detail of its context in Table \ref{tab:pi_experiment}.

\begin{table}[h!]
\centering
\begin{tabular}{p{1cm} p{3cm} p{7cm}}
\hline
\textbf{Line\#} & \textbf{Pi output} & \textbf{Notes} \\
\hline
1 & {\fontsize{8pt}{\baselineskip}\selectfont\texttt{cameron@cryp...}} & Experiment run with 3,200 iterations of ML-KEM-1024: `python ./experimenter.py \text{-}\text{-}algorithm ML-KEM-1024 \text{-}\text{-}iterations 3200' \\
2 & {\fontsize{8pt}{\baselineskip}\selectfont\texttt{UDP: Sent 'GET...}} & This notes that a `GETREADY' has been sent to the Collector PC with parameters\\
3 & {\fontsize{8pt}{\baselineskip}\selectfont\texttt{Starting exp...}} & Notes the experiment with this ID is starting - informing user\\
4 & {\fontsize{8pt}{\baselineskip}\selectfont\texttt{Setting up en...}} & Notes fan and CPU are being set to static values for test\\
5 & {\fontsize{8pt}{\baselineskip}\selectfont\texttt{Start Tempe...}} & It notes the temperature recorded by the Pi's CPU prior to the experiment\\
6 & {\fontsize{8pt}{\baselineskip}\selectfont\texttt{UDP: Sent 'STA...}} & The START message is sent to the Collector so it starts recording energy information\\
7 & {\fontsize{8pt}{\baselineskip}\selectfont\texttt{STARTing exp...}} & Immediately afterwards it starts the experiment - 3200 lots of KEM-1024\\
8 & {\fontsize{8pt}{\baselineskip}\selectfont\texttt{Experiment f...}} & No output until this message after Pi experiment finishes\\
9 & {\fontsize{8pt}{\baselineskip}\selectfont\texttt{UDP: Sent 'STO...}} & A STOP message is set to collector to cease energy recordings\\
10 & {\fontsize{8pt}{\baselineskip}\selectfont\texttt{Time to run ...}} & Notes Wall-clock time for the experiment\\
11 & {\fontsize{8pt}{\baselineskip}\selectfont\texttt{Start Tempe...}} & Shows start temperature of Pi\\
12 & {\fontsize{8pt}{\baselineskip}\selectfont\texttt{Stop Temper...}} & Displays temperature of Pi after experiment has completed\\
13 & {\fontsize{8pt}{\baselineskip}\selectfont\texttt{Environment ...}} & Notes that the fan and CPU have been set back to defaults at end of experiment\\
14 & {\fontsize{8pt}{\baselineskip}\selectfont\texttt{cameron@cryp...}} & Command line as experiment complete. \\
\hline
\end{tabular}
\caption{Pi Experiment Output Explainer for Figure \ref{PiSingle}}
\label{tab:pi_experiment}
\end{table}

\begin{figure}[h!]
  \centering
  \includegraphics[width=12cm]{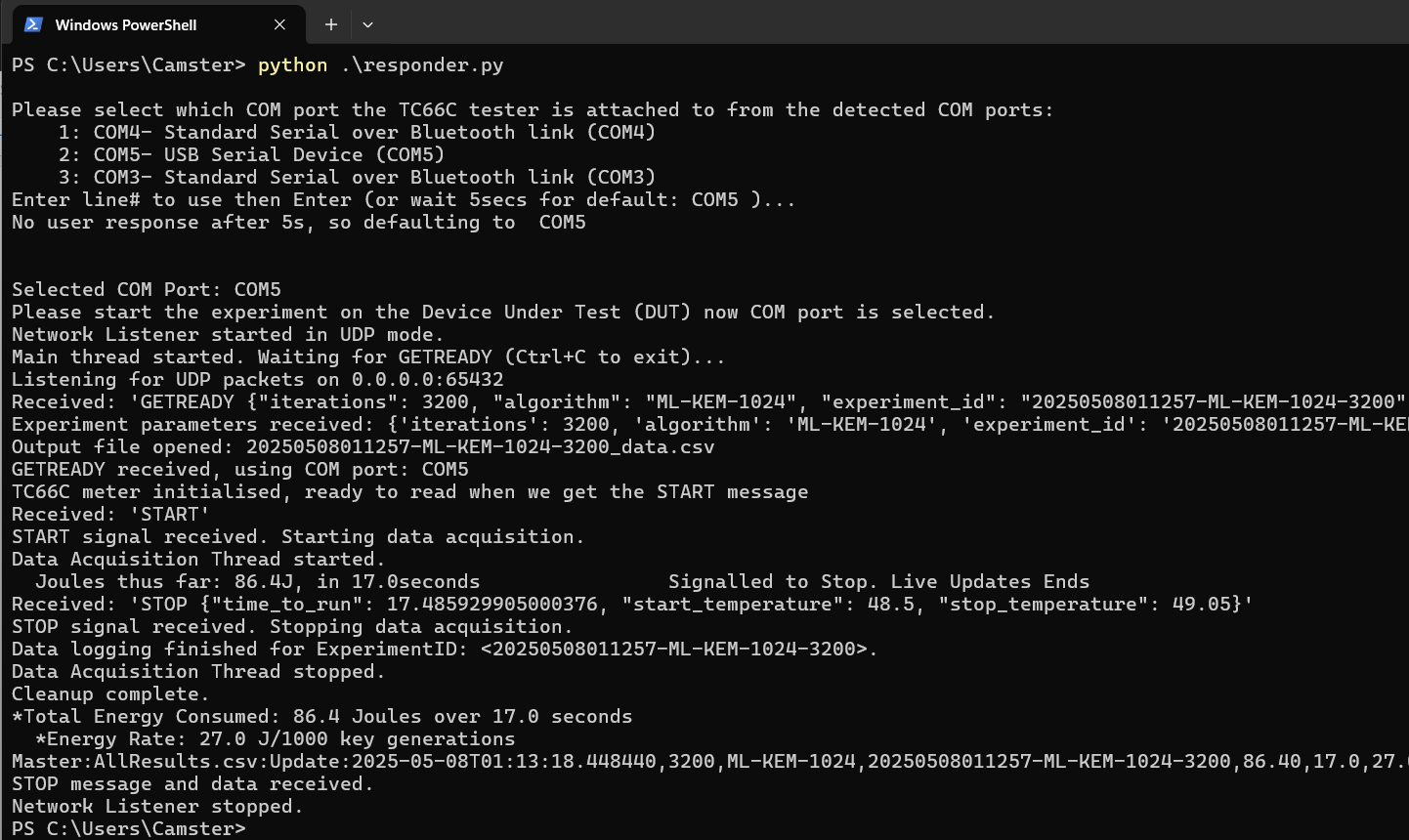}
  \caption{Command Line from the Windows Collector - the `Joules this far' line is updated several times per second to keep the user updated with the status and confirm that all is running. This is the corresponding output from the same experiment as per Figure \ref{PiSingle} above. An explainer can be found in Table \ref{tab:win_experiment} below.}
  \label{WinSingle}
\end{figure}

The collector device should be set up in advance of starting the experiment to receive its output.
The corresponding interaction to the 3200 count ML-KEM-1024 experiment is seen from the Collector device's perspective in the Figure \ref{WinSingle} screen capture, together with a line-by-line explanation in Table \ref{tab:win_experiment} which follows.

\begin{table}[h!]
\centering
\begin{tabular}{p{1cm} p{3cm} p{7cm}}
\hline
\textbf{Line\#} & \textbf{Win output} & \textbf{Notes} \\
\hline
1 & {\fontsize{8pt}{\baselineskip}\selectfont\texttt{PS C:Users:Cams...}} & Run the command to start the Collection\\
2 & {\fontsize{8pt}{\baselineskip}\selectfont\texttt{Please select ...}} & (can force/bypass selection e.g. \text{-}\text{-}com COM5)\\
3 & {\fontsize{8pt}{\baselineskip}\selectfont\texttt{1: COM4.3:COM3...}} & shows detected COM ports on the PC if not provided\\
4 & {\fontsize{8pt}{\baselineskip}\selectfont\texttt{Enter line\# ...}} & User can select the COM line\\
5 & {\fontsize{8pt}{\baselineskip}\selectfont\texttt{No user respo...}} & It defaults if response times out\\
6 & {\fontsize{8pt}{\baselineskip}\selectfont\texttt{Selected COM ...}} & Details which port is in use\\
7 & {\fontsize{8pt}{\baselineskip}\selectfont\texttt{Please start ...}} & Note that Collector is ready to receive START message\\
8 & {\fontsize{8pt}{\baselineskip}\selectfont\texttt{Network Liste...}} & User notification that UDP will be used\\
9 & {\fontsize{8pt}{\baselineskip}\selectfont\texttt{Main thread s...}} & and waiting for a GETREADY from Pi\\
12 & {\fontsize{8pt}{\baselineskip}\selectfont\texttt{Listening for ...}} & All interfaces listening on port\\
13 & {\fontsize{8pt}{\baselineskip}\selectfont\texttt{Received: 'GET...}} & GETREADY received from experiment and Pi\\
14 & {\fontsize{8pt}{\baselineskip}\selectfont\texttt{Experiment pa...}} & Prints the parameters of experiment received from Pi in packet\\
15 & {\fontsize{8pt}{\baselineskip}\selectfont\texttt{Output file o...}} & Specific Experiment logfile opened and ready\\
16 & {\fontsize{8pt}{\baselineskip}\selectfont\texttt{GETREADY rece...}} & notes that COM5 serial port will be used\\
17 & {\fontsize{8pt}{\baselineskip}\selectfont\texttt{TC66C meter i...}} & initialises thread for data coming in from meter on this COM\\
18 & {\fontsize{8pt}{\baselineskip}\selectfont\texttt{Received: 'STA...}} & Network START received on network thread\\
19 & {\fontsize{8pt}{\baselineskip}\selectfont\texttt{START signal ...}} & Message user to say we're getting started\\
20 & {\fontsize{8pt}{\baselineskip}\selectfont\texttt{Data Acquisiti...}} & Starts getting data from TC66C meter, counters to 0\\
21 & {\fontsize{8pt}{\baselineskip}\selectfont\texttt{Joules thus f...}} & This line auto-updates - usually a few times a second, shows cumulative totals of energy and time elapsed in experiment\\
22 & {\fontsize{8pt}{\baselineskip}\selectfont\texttt{Received: 'STO...}} & Experiment has ended - Pi has sent STOP\\
23 & {\fontsize{8pt}{\baselineskip}\selectfont\texttt{STOP signal r...}} & Stop recording energy data, the totals are known now\\
24 & {\fontsize{8pt}{\baselineskip}\selectfont\texttt{Data logging f...}} & Experiment log file flushed and saved\\
25 & {\fontsize{8pt}{\baselineskip}\selectfont\texttt{Data Acquisiti...}} & Close down the connection to meter as no more data to come\\
26 & {\fontsize{8pt}{\baselineskip}\selectfont\texttt{Cleanup comple...}} & Verifying network, meter threads and individual log file closed\\
27 & {\fontsize{8pt}{\baselineskip}\selectfont\texttt{*Total Energy ...}} & Prints a summary of the result totals\\
28 & {\fontsize{8pt}{\baselineskip}\selectfont\texttt{*Energy Rate:...}} & Prints the energy rate per 1,000 keys generated for experiment\\
29 & {\fontsize{8pt}{\baselineskip}\selectfont\texttt{Master:AllResu...}} & This is written as a one-liner to the master log AllResults.csv\\
30 & {\fontsize{8pt}{\baselineskip}\selectfont\texttt{STOP message ...}} & Flushing the threads - this message is last out\\
31 & {\fontsize{8pt}{\baselineskip}\selectfont\texttt{Network Liste...}} & No longer listening to the network, we're done\\
32 & {\fontsize{8pt}{\baselineskip}\selectfont\texttt{PS C::Users:Cams...}} & Returns to the command line\\
\hline
\end{tabular}
\caption{Windows Measurement Output Explainer for Figure \ref{WinSingle}}
\label{tab:win_experiment}
\end{table}

\subsubsection{Batch Run}
In the GitHub project repository, sample input files are provided which are used by the 'batch\_ experimenter.py` script.  
The files have two comma separated values, the first of which is the key generation algorithm parameters conveyed to OpenSSL, and the second is the desired number of iterations of the test.  
The higher the number of iterations the longer the experiment takes, and the better accuracy that should be obtained for the energy rate output (J/1,000 key generations). 

However, as outlined in the results plot (Figure \ref{TimeComparePlot}), generating an RSA key with a `large' key size takes significantly longer, so the iterations for these are typically scaled down in the input source file.
This example illustrates that for this 100,000 iteration experiment, only 200 RSA-4096 keys are generated as the process is orders-of-magnitude slower, so this is scaled to a feasible number which should be a similar order of magnitude for the elapsed time.

\begin{itemize}
    \item \texttt{EC -pkeyopt ec\_paramgen\_curve:secp160r1,100000}
    \item \texttt{EC -pkeyopt ec\_paramgen\_curve:P-521,100000}
    \item \texttt{NULL,100000}
    \item \texttt{ML-KEM-1024,100000}
    \item \texttt{RSA -pkeyopt rsa\_keygen\_bits:4096,200}
\end{itemize}

\noindent For testing it is possible to override the number of iterations via the command line with the optional \text{ -}\text{-}iterations argument.  
An example is: 

`python ./batch\_experimenter.py \text{ -}\text{-}iterations 10 100kSourceRSAscaled.txt'  where instead of 100k for some of the options, actually only 10 key generations would be carried out for each algorithm-type in the source file.
Many of the pieces of code here have command line options. To find out more, details are in the comment block at the top of each piece of code, or by adding the `\text{-}\text{-}help' option as an argument when running the file.

\subsubsection{Results Files}
Finally in this Appendix, the output files found in the repository are described:
For each experiment (there may be multiple experiments in a batch), an individual .csv file is generated, which has all data from every poll of the meter.  
These files are named e.g. `20250507133228-ML-KEM-1024-3200\_data.csv' which is sortable and designates the date and time, the algorithm and the iterations of the experiment.
Additionally a log file `AllResults.csv' has one line of results added for each completed experiment.
This includes the timestamp, algorithm, iteration count as well as information on both energy and time rates and totals. 
This ensures that the results of all experiments then can be recovered at a later time, and as required, even if results have already cleared off the console.

More information about the code, input and output files can be found in the readmes in the GitHub repository \cite{CPGITHUB} for the project, or in the comment blocks of the Python files built to support the energy recording framework.




\bibliographystyle{IEEEtran}
\bibliography{main}

\end{document}